\documentclass[
reprint,
superscriptaddress,
 amsmath,amssymb,
 aps,
prb,
]{revtex4-2}

\usepackage{graphicx}
\usepackage{dcolumn}
\usepackage{bm}
\usepackage{hyperref}
\usepackage{upgreek}

\begin{document}

\preprint{APS/123-QED}

\title{Quench of the electronic order in a strongly-coupled charge-density-wave system by enhanced lattice fluctuations}

\author{Manuel Tuniz}
\email{manuel.tuniz@phd.units.it}
\affiliation{Dipartimento di Fisica, Università degli Studi di Trieste, 34127 Trieste, Italy}
\affiliation{CNR - Istituto Officina dei Materiali (IOM), Unità di Trieste, Strada Statale 14, km 163.5, 34149 Basovizza (TS), Italy}
 
\author{Denny Puntel}%
\affiliation{Dipartimento di Fisica, Università degli Studi di Trieste, 34127 Trieste, Italy}
 
\author{Wibke Bronsch}%
\affiliation{Elettra - Sincrotrone Trieste S.C.p.A., Strada Statale 14, km 163.5, Trieste, Italy}
  
\author{Francesco Sammartino}%
\affiliation{Dipartimento di Fisica, Università degli Studi di Trieste, 34127 Trieste, Italy}

\author{Gian Marco Pierantozzi}
\affiliation{CNR - Istituto Officina dei Materiali (IOM), Unità di Trieste, Strada Statale 14, km 163.5, 34149 Basovizza (TS), Italy}

\author{Riccardo Cucini}
\affiliation{CNR - Istituto Officina dei Materiali (IOM), Unità di Trieste, Strada Statale 14, km 163.5, 34149 Basovizza (TS), Italy}

\author{Fulvio Parmigiani}
\affiliation{Dipartimento di Fisica, Università degli Studi di Trieste, 34127 Trieste, Italy}
\affiliation{Elettra - Sincrotrone Trieste S.C.p.A., Strada Statale 14, km 163.5, Trieste, Italy}
\affiliation{International Faculty, University of Cologne, Albertus-Magnus-Platz, 50923 Cologne, Germany}

 \author{Federico Cilento}%
 \email{federico.cilento@elettra.eu}
 \affiliation{Elettra - Sincrotrone Trieste S.C.p.A., Strada Statale 14, km 163.5, Trieste, Italy}

\date{\today}

\begin{abstract}

	Charge-density-wave (CDW) materials having a strong electron-phonon coupling provide a powerful platform for investigating the intricate interplay between lattice fluctuations and a macroscopic quantum order. Using time- and angle-resolved photoemission spectroscopy (TR-ARPES), we reveal that the CDW gap closure in VTe$_2$ is dominated by an incoherent process evolving on a sub-picosecond timescale, challenging the conventional view that the gap dynamics is primarily governed by the excitation of the CDW amplitude modes. Our findings, supported by a three-temperature model, show that the CDW gap evolution can be described by considering the population of a subset of strongly-coupled optical phonon modes, which leads to an increase in the lattice fluctuations. This microscopic framework extends beyond VTe$_2$, offering a universal perspective for understanding the light-induced phase transition in strongly-coupled CDW systems.

\end{abstract}

\maketitle
	
	\section{Introduction}

	Among the various microscopic interactions shaping quantum materials, electron-phonon coupling (EPC) has been a persistent subject of study since it stands at the origin of a broad variety of phenomena. One of the manifestations of strong EPC, which is lately attracting a large interest, is the appearance of a charge-density wave (CDW) phase in many materials \cite{Rossnagel_2011, Kang_2022, Comin_2014}. In CDW systems, electrons and phonons cooperatively interact to form a new symmetry-broken state below a critical transition temperature \cite{Gruner_1994}. The resulting low-temperature phase is characterized by the coexistence of a spatial modulation of the electron density and a periodic distortion of the crystalline structure \cite{Gruner_1994}. While the charge-density-wave transition has been traditionally described in the weak EPC limit \cite{Gruner_1988}, the discovery of compounds exhibiting strong momentum-dependent EPC has motivated both theoretical \cite{McMillan_1977, Varma_1983, Inglesfield_1980, Zhu_2015} and experimental studies \cite{Tediosi_2009, Hofmann_2019, Cao_2023}. Due to the complexity of these systems, however, a unified microscopic understanding of the CDW transition and its dynamics remains elusive \cite{Rossnagel_2011}. 
	
	Giving the pivotal role played by the EPC in determining the CDW phase transition, an increase of its strength leads to a dramatic deviation from the conventional Peierls picture of the CDW transition \cite{Gruner_1994, Peierls_1955}. Indeed, while in the weak-coupling limit the transition from the CDW to the normal phase is driven by the electronic entropy, in the strong-coupling limit the phase transition is instead driven by the lattice entropy, thus resulting in an order-disorder transition \cite{McMillan_1977, Tosatti_1995}. As a result, strongly-coupled CDW materials provide a compelling platform to study the interplay between lattice fluctuations and the emergence of a macroscopic quantum order. Furthermore, controlling the CDW phase and understanding its ultrafast dynamics are essential steps toward the development of next-generation electronic devices based on these materials \cite{Vaskivskyi_2019}. More broadly, lattice fluctuations are not only key to understand the CDW dynamics in VTe$_2$, but they are also believed to play a crucial role in the phase transition of a wide range of quantum materials \cite{Fechner_2024, Wall_2018}.\looseness-1

\begin{figure*}[t!]
\includegraphics [width=0.93\textwidth ]{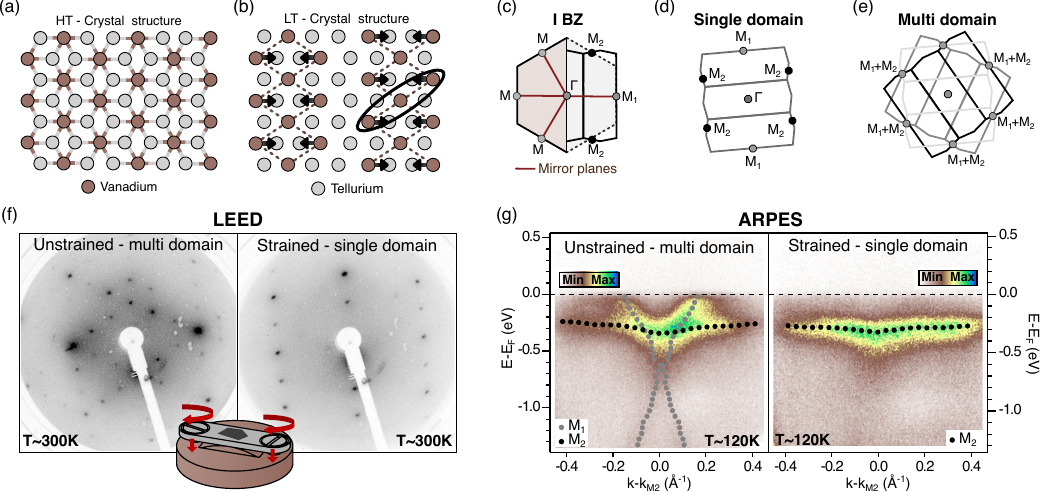}
\caption{(a) Projection of an high-temperature (HT) single VTe$_2$ layer on the pseudo-hexagonal plane. (b) Projection of a low-temperature (LT) single VTe$_2$ layer on the pseudo-hexagonal plane. The black arrows show the displacement direction of the vanadium atoms, while the black ellipse and dashed lines highlight the formation of a trimer-like bonding between three vanadium atoms, resulting in a double zigzag structure. (c) High-temperature (left) and low-temperature (right) first Brillouin zones of VTe$_2$. The mirror planes of the two BZs are also shown. (d) View of the reciprocal space as seen by a probe beam smaller than the size of the domains. (e) View of the reciprocal space as seen by a probe beam larger than the domain size. Due to the averaging over different domains, the M$_1$ and M$_2$ points are perfectly overlapped. (f) LEED images taken before (left panel) and after (right panel) the application of strain on the same VTe$_2$ sample. A single domain with $3 \times 1$ superstructure is observed in the strained case. The images have been taken with a kinetic energy of $75\,$eV. The inset depicts the strain device used in the experiments. (g) ARPES spectra acquired along the K-M$_2$-K direction on an unstrained (left) and on a strained (right) samples.\looseness-1}
\label{fig:1}
\end{figure*}

	In this work, studying the CDW gap dynamics by means of TR-ARPES experiments, we demonstrate that in the strongly-coupled CDW compound VTe$_{2}$, the light-induced phase transition is determined by an incoherent process that evolves on a timescale much slower than the one expected for a conventional displacive transition \cite{Wall_2018}. By implementing a three-temperature model, we demonstrate that the CDW gap dynamics can be described considering the population of a subset of strongly-coupled optical phonon modes, which determine an increase of the lattice fluctuations, and hence of the transient disorder of the system.

	\section{Methods}
	
	High-quality VTe$_2$ single crystals were provided by HQ Graphene. The samples were cleaved \emph{in situ}, at a base pressure better than $5\times 10^{-10}\,$mbar and at room temperature. ARPES and TR-ARPES experiments were performed using a state-of-the-art HHG source equipped with a time-preserving monochromator and synchronized to an OPA system delivering the pump pulses \cite{Cucini_2020}. For the present experiments, the pump and probe photon energies were set to $1.77\,$ and $21.6\,$eV respectively. The ultimate energy and time resolutions were $\sim 55\,$meV and $\sim 110\,$fs, respectively, while the FWHM of the probe beam at the sample position was of the order of $100\, \upmu$m. An home-built strain device was mounted on the sample holder, to perform photoemission experiments on strained samples (further details are reported in the SM).

	\section{Experimental Results}

	\begin{figure*}[t!]
	\includegraphics [width=0.85\textwidth]{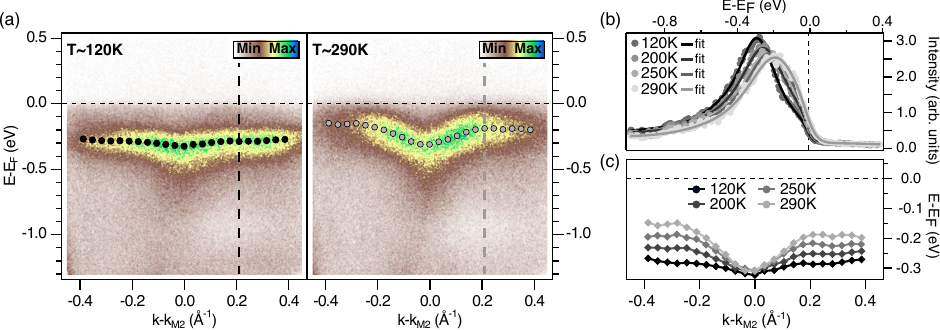}
	\caption{(a) Photoemission spectra acquired along the K-M$_2$-K direction at $120\,$K and $290\,$K. The grey circles denote the central position of the Lorentzian peaks used to fit the EDCs, hence they show the dispersion of the vanadium 3d band. (b) EDCs extracted from the momentum region denoted by the dashed lines in (a) for four temperatures. (c) Evolution of the dispersion of the hybridized vanadium 3d band as a function of the temperature.} 
	\label{fig:2}
	\end{figure*}

	\subsection*{Strain-induced domain stabilization} 
		
	In the transition-metal dichalcogenide compound VTe$_2$, the presence of a strong EPC results in the opening of a CDW gap  whose size is five times larger than the one predicted by the weak-coupling limit, making this compound an ideal candidate for studying the effect of enhanced lattice fluctuations on the  CDW phase \cite{Gruner_1994}. However, the study of the gap dynamics is hindered by the presence of multiple CDW domains that characterize the low-temperature phase. Indeed, at $480\,$K the system undergoes a first order phase transition towards a new commensurate $3 \times 1 \times 3$ charge-ordered phase (Fig.~\ref{fig:1}(a) and (b)), that leads to the formation of three equivalent in-plane domains, rotated by $120$ degrees \cite{Bronsema_1984, Ohtani_1981, Vinokurov_2009}. As depicted in Fig.~\ref{fig:1}(d) and (e), when a probe beam larger than the typical domain size is used, the resulting superposition of BZs causes the M$_1$ and M$_2$ points to overlap, thereby hindering the study of the CDW gap that opens at the M$_2$ point. To overcome this issue, following an emerging approach to control the domain structure of complex materials \cite{Nicholson_2021}, we developed a strain device capable of delivering uniaxial tensile strain to the VTe$_2$ crystals. By breaking the equivalence between the different domains, the application of strain results in the formation of larger domains. The effectiveness of this approach was assessed by means of low energy electron diffraction (LEED) and ARPES experiments. Figure \ref{fig:1}(f) shows a comparison between two LEED images acquired on the same sample before (left) and after (right) the application of a tensile strain of $\sim 0.2\,\%$ (details on the strain application and its quantification are reported in the SM). The effect of the applied strain is clearly visible from these images. While the complex pattern that appears in the unstrained sample arises from the superposition of the three different domains, the one observed on the strained sample clearly denotes the presence of a single domain in the probed region. To further confirm the effectiveness of our approach, in Fig.~\ref{fig:1}(g) we report a comparison between two ARPES spectra acquired on an unstrained (left) and on a strained (right) samples. As for the LEED image, the ARPES map acquired on the unstrained sample shows a superposition of bands coming from the M$_1$ and the M$_2$ points of different Brillouin zones (BZs) \cite{Mitsuishi_2020}. In contrast, the ARPES map acquired on the strained sample only shows the nearly-flat band from the M$_2$ point. These spectra therefore demonstrate the possibility of performing photoemission experiments on a single domain by applying a tensile strain to the VTe$_{2}$ crystals.  Since the only detectable effect of the applied strain is the promotion of larger domains, all the ARPES experiments presented below were performed on strained samples.

	\subsection*{ARPES measurements}
  
	We start by analyzing the effect of a temperature change on the CDW gap that opens at M$_2$. Figure \ref{fig:2}(a) shows the band structure measured along the K-M$_2$-K direction, at $120$ and $290\,$K. As shown in Fig.~\ref{fig:1}(c), in the normal phase, the M points reside on a mirror plane which prohibits the hybridization between the Te 5p orbitals (odd with respect to the mirror plane) and the V 3d orbitals (even with respect to the mirror plane). In the low temperature phase, instead, while the M$_1$ points keep a similar situation with respect to the high-temperature phase, the M$_2$ points do not reside anymore on a mirror plane. Hence the V 3d and the Te 5p orbitals can now mix, leading to a significant change in the band structure \cite{Mitsuishi_2020}. The V-shaped band that characterizes the M points in the normal phase of VTe$_{2}$ is replaced by a weakly-dispersing band at E$_{\text{B}}\sim 0.25\,$eV, whose flatness reflects the localized nature of the d$_{\text{yz}}$ and d$_{\text{zx}}$ orbitals in real space \cite{Mitsuishi_2020}. As expected from the change in the lattice symmetry, by increasing the temperature, we observe a folding of the hybridized vanadium 3d band towards the Fermi level, in such a way that the V-shaped band is partially restored. Figures \ref{fig:2}(b) and (c) show the evolution of the position of the vanadium band (details on the fitting procedure are reported in the SM) as a function of temperature. Despite the presence of a first order phase transition, far from the critical temperature we observe a progressive folding of the vanadium band, with the consequent partial closure of the CDW gap.

	\begin{figure*}[t!]
	\includegraphics[width=\textwidth ]{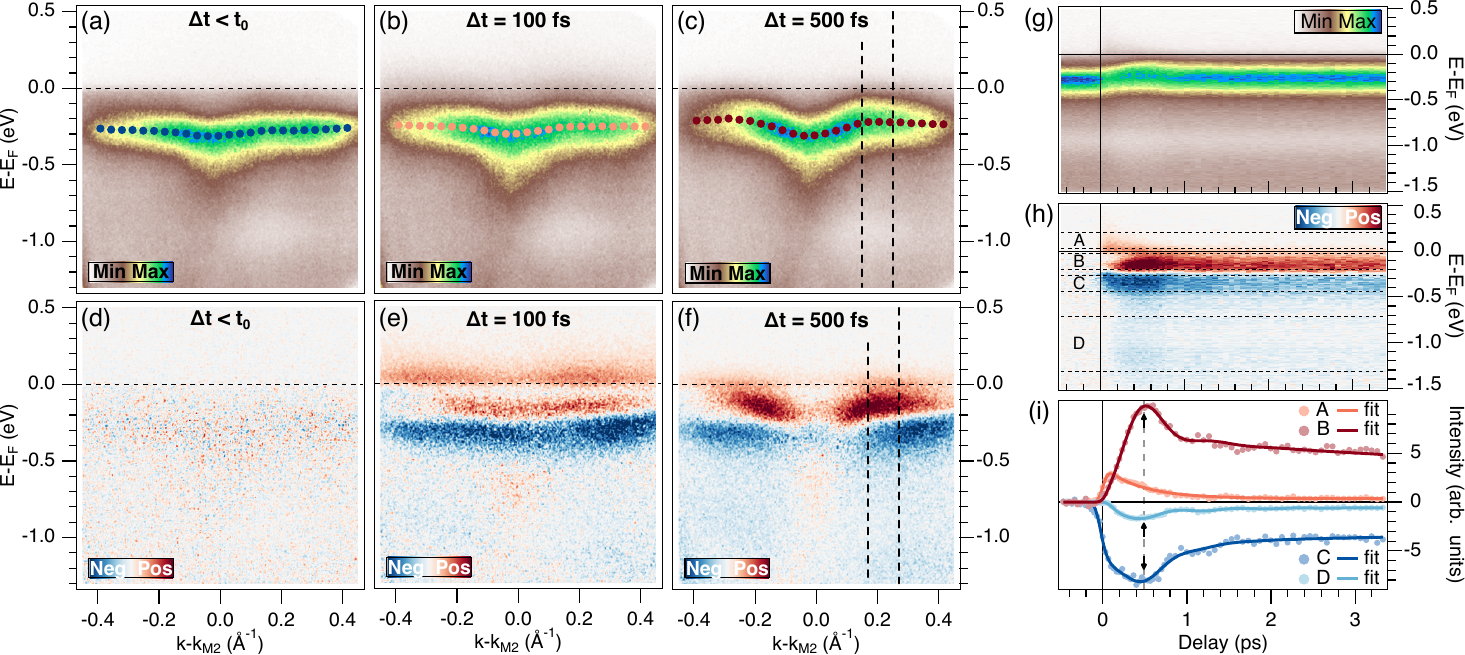}
	\caption{(a)-(c) Selected ARPES spectra acquired along the K-M$_2$-K direction at different pump-probe delays: (a) before the arrival of the pump, (b) and (c) respectively $100$ and $500\,$fs after the arrival of the pump pulse. The colored circles denote the central position of the vanadium band extracted from the fits of the EDCs. (d)-(f) Differential ARPES maps showing the changes in the photoemission intensity induced by the pump pulse. (g) and (h)  EDCs and dEDCSs extracted by integrating the photoemission intensity from the momentum region delimited by the dashed lines in (c) and (f), plotted as a function of the pump-probe delay. (i) Evolution of the photoemission intensity extracted from the boxes reported in (h). The gray dashed line shows the pump-probe delay at which traces B and C reach their maximum change. The measurements were performed at a base temperature of $\sim 120\,$K, and using an absorbed fluence of $\sim 1.2\,$mJ/cm$^2$.} 
	\label{fig:3}
	\end{figure*}

	\subsection*{TR-ARPES measurements}

	Having characterized the temperature-driven phase transition, we now investigate the possibility to melt the CDW order using an ultrashort pump pulse ($h\nu_{\text{pump}}=1.77\,$eV). Figure \ref{fig:3} shows three ARPES spectra acquired at different pump-probe delays, along with the differential maps that highlight the changes induced by the pump pulse. Panels (a)-(c) show that, after photoexcitation, the dispersion of the vanadium band is modified in a way similar to the one observed in the temperature-driven phase transition. In particular, we observe a partial reappearance of the V-shaped band that characterizes the normal phase of VTe$_{2}$. Interestingly, from the comparison of the three spectra it appears that the changes in the dispersion of the vanadium 3d band are more pronounced in the spectrum acquired at a pump-probe delay of $500\,$fs than in the one at $100\,$fs. These differences are well resolved in the differential images shown in Fig.~\ref{fig:3}(e) and (f), where two qualitatively different scenarios appear. While at $100\,$fs (panel (e)) only the fingerprint of a small shift of the whole vanadium band towards the Fermi level is detected, the pattern visible in panel (f) is likely to arise from the folding of the vanadium band. To study the full dynamics of the vanadium band, in Fig.~\ref{fig:3}(g) and (h) we report the evolution of the EDCs and the differential EDCs (dEDCs), integrated in the momentum region delimited by the dashed lines shown in Fig.~\ref{fig:3}(c) and (f), as a function of the pump-probe delay. Having verified that the dynamics extracted from the two sides (left and right with respect to k-k$_{\text{M2}} = 0$) of the vanadium band are equivalent, here only the ones extracted from the right side of the V-shaped band are analyzed. Figure \ref{fig:3}(i) shows the traces obtained by integrating the photoemission intensity inside the boxes drawn in panel (h). While trace A shows a rise time limited only by the time resolution of the setup, and reaches its maximum about $\sim 100\,$fs after the arrival of the pump pulse, the other traces show larger rise times. In particular, the maximum change for traces B and C is recorded at a pump-probe delay of $\sim 500\,$fs, indicating that the folding of the vanadium band, and the consequent closure of the CDW gap, evolves on a much slower timescale with respect to the one of the hot carriers described by trace A. Moreover, the absence of marked oscillations in the gap size signals that the photoinduced phase transition in VTe$_{2}$ is not driven by the excitation of the CDW amplitude modes, as instead observed in many other CDW systems. Indeed, in those systems, the maximum closure of the CDW gap occurs on a timescale corresponding to half period of the CDW amplitude mode \cite{Sohrt_2014, Yang_2020, Schmitt_2008, Rettig_2016, Schmitt_2011, Maklar_2023}. As demonstrated by Tuniz \emph{et al.} \cite{Tuniz_2023RR}, VTe$_2$ has two amplitude modes with a low-temperature frequency of $1.6\,$THz and $2.5\,$THz, for which a characteristic timescale of $300$ and $200\,$fs would be expected, respectively. The fact that the timescales observed here are almost two times longer, suggests that the melting of the charge-ordered is not dominated by a conventional displacive process.

	\begin{figure*}[t!]
	\includegraphics[width=0.9\textwidth ]{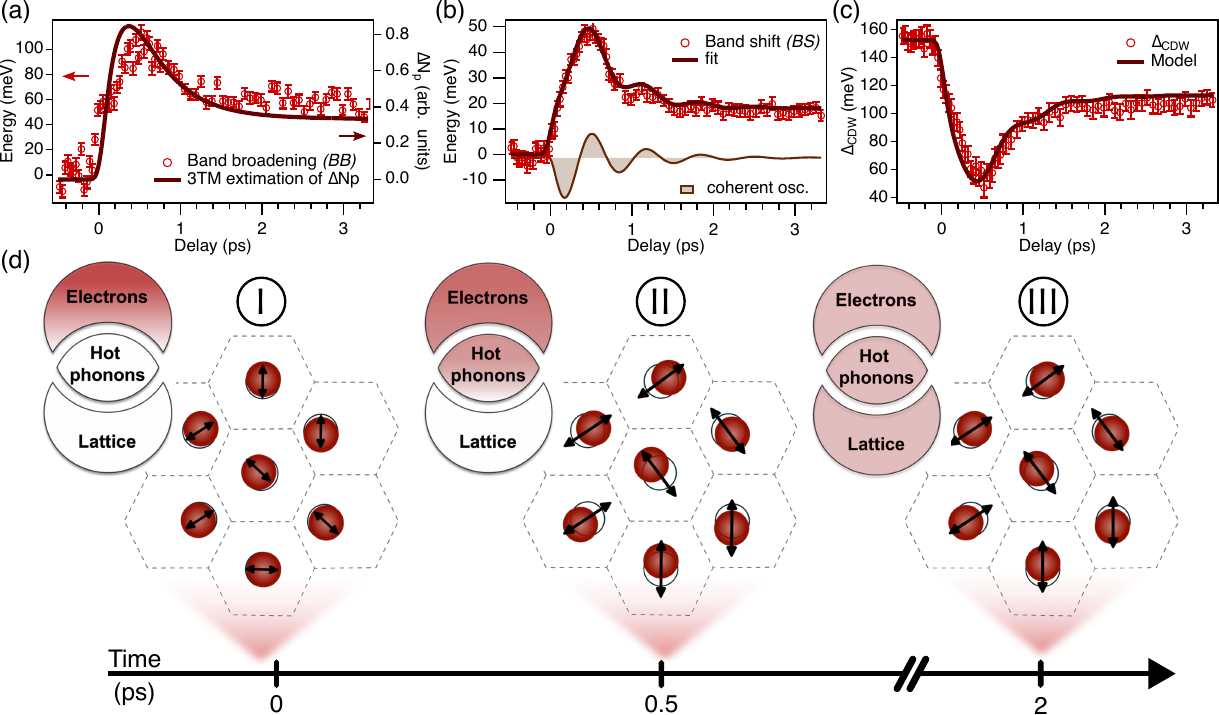}
	\caption{(a) Comparison between the transient broadening of the CDW gap and the change in the population of the strongly-coupled optical phonon modes. (b) Fit of the CDW gap closure obtained by summing to the trace describing the transient change in the population of the strongly-coupled modes (rescaled by a constant) an oscillating term with the frequency of the CDW amplitude mode having the lowest energy. (c) Comparison between the dynamics of the CDW gap (distance between the leading edge of the lower branch and the Fermi level) and the model obtained by summing the curves shown in (a) and (b). (d) Sketch showing the partial loss of the long-range CDW order, occurring as a consequence of the incoherent excitation of the strongly-coupled optical phonon modes. The red spheres represent the vanadium atoms while the black arrows depict the amplitude of their displacements. (I) Before the arrival of the pump pulse the three subsystems are in equilibrium (T$_{\text{e}}$ = T$_{\text{p}}$ = T$_{\text{e}}$ $\ll$ T$_{\textsc{cdw}}$) and the system is characterized by a well-defined CDW order. The energy injected by the pump in the system is initially absorbed by the electrons and then transferred to a subset of strongly-coupled optical phonon modes. (II) The excitation of these phonon modes, without a macroscopic phase coherence, leads to a partial loss of the CDW long-range order of the system. (III) After $\approx 2\,$ps from the arrival of the pump pulse, the three subsystems are again in equilibrium and from there on the relaxation dynamics is governed solely by heat diffusion.}
	\label{fig:5}
	\end{figure*}

	 In general, a solid-solid phase transition, where the symmetry of the crystal is raised, can be determined by either a displacive process or an order-disorder process \cite{Wall_2018, Budai_2014, Huber_2014}. While in a displacive transition the atoms collaboratively reshuffle their positions under the effect of a few well-defined spatially coherent vibrational modes, in an order-disorder transition the atoms move from the low to the high-symmetry structure in such a way that the motion remains correlated only along one direction, while along the others it proceeds in a spatially incoherent manner, with no characteristic correlation length \cite{Wall_2018, Bruce_1980, Lindenberg_2005, delaPenaMunoz_2023, Keen_2015, Johnson_2024}. In the present case of VTe$_2$, the emerging picture is that the gap dynamics is determined by a loss of the long-range CDW order triggered by an increase of lattice fluctuations. We emphasize that, in the present case, the role played by the increased lattice fluctuations can be isolated only thanks to the large energy gaps that characterize the strongly-coupled CDW systems like VTe$_2$, which make the electronic entropy (that otherwise would have overshadowed these effects) unimportant in determining the phase transition \cite{McMillan_1977}.
	 
	  In the light-induced process considered, the increase in the lattice fluctuations occurs as a consequence of the energy transfer from the excited electrons to the lattice \cite{Rameau_2016, Cotret_2019, Girault_1989}. Therefore, to model this energy flow, we implement a three-temperature model ($3$TM) and we compare the results obtained to the ones derived from a line shape analysis of the CDW gap (details on the fitting procedure are reported in the SM) \cite{Perfetti_2007}. Specifically, by fitting the dynamics of the electronic temperature extracted at M$_1$, we can derive the temperature evolution of the optical phonon modes which are more strongly coupled to the electronic system. These simulations were performed assuming the phonon branch at $\omega_{\text{p}}\sim 24\,$meV to be the one more strongly coupled, with a resulting e-ph coupling constant of $\lambda = 0.32$ (a detailed discussion is reported in the SM).  Starting from the evolution of the temperature ($T_{\text{p}}$) of these strongly-coupled modes, and assuming that their population ($n_{\text{p}}$) can be described by the Bose-Einstein distribution:\looseness-1
	\begin{equation}
		n_{\text{p}} (t) \propto \left[ \exp \left( \frac{\hbar \omega_{\text{p}}}{\text{k}_{\text{B}} T_{\text{p}} (t)}\right) -1 \right]^{-1},
	\end{equation}

	it is possible to calculate the change in the population of the strongly-coupled modes due to the increase of their temperature, triggered by the hot carrier relaxation \cite{Cotret_2019}. We emphasize that these modes are excited without a macroscopic phase coherence, and therefore their excitation leads to an increase of the lattice fluctuations in the system.  Increased lattice fluctuations give rise to a smearing of the gap edges \cite{Degiorgi_1994, McKenzie_1992}, therefore establishing a link between the transient disorder in the system and the broadening of the CDW gap \cite{Kwok_1990}.
	
	We proceed by comparing the timescales over which these processes evolve. Figure \ref{fig:5}(a) shows a comparison between the transient broadening ($BB(t)$) of the CDW gap and the change in the population of the strongly-coupled modes ($\Delta n_{\text{p}}(t)$). Interestingly, these two traces evolve on similar timescales. The excitation of the strongly-coupled optical phonon modes is governed by the hot carrier relaxation through the electron-phonon coupling constant, therefore the maximum change in the population of the modes is achieved only $400-500\,$fs after the arrival of the pump pulse. This timescale matches well the one observed for the smearing of the CDW gap. Moreover, also the second relaxation process of both traces is similar, with the two showing a slow recovery dynamics for pump-probe delays larger than $2\,$ps. 
	
	The results obtained by applying the $3$TM can be used to model also the dynamics of the gap closure. Indeed, the broadening of the gap discussed above entails a reduction of the gap size itself, and therefore a reduction of the energy gain of the electronic subsystem with the consequent reduction of the amplitude of the lattice distortion. With this in mind we can compare the incoherent dynamics of the closure of the CDW gap with the one describing the change in the population of the strongly coupled optical phonon modes (rescaled by a constant factor). Moreover, given the presence of small bumps in the trace showing the evolution of the band shift, on top of this incoherent dynamics we summed a damped cosine oscillation with the frequency of the lowest-energy CDW amplitude mode \cite{Tuniz_2023RR}. Therefore, the evolution of the band shift ($BS$) has been modeled using the equation:\looseness-1
	\begin{equation}
		BS(t) = A \cdot \Delta n_{\text{p}}(t)+ I \, e^{-t/\tau} \cos(\omega_{\textsc{am}}t + \phi),
	\label{eq:fit_BS}
	\end{equation}

	where $A$ is the constant used to rescale the change in the population of the strongly-coupled modes. $I$ is the intensity of the oscillating component  with angular frequency $\omega_{\textsc{am}}$, decay time $\tau$ and phase $\phi$. The result of the fit obtained by using Eq.~\ref{eq:fit_BS} is shown in Fig.~\ref{fig:5}(b), together with the coherent  part of the fit, reported separately. As for the case of the gap broadening, the timescales extracted from the $3$TM match well with the ones extracted from the experiments, hence corroborating the thermal origin of the dynamics observed. 
	
	Finally, being able to describe the dynamics of both gap closure and gap broadening in terms of the change in the population of the strongly-coupled modes, it is possible to sum together the two models to  describe the dynamics of the CDW gap: 
	\begin{equation}
		\Delta_{\textsc{cdw}}(t) = \Delta_{\textsc{cdw}}(0) - BS(t)- \frac{1}{2} BB(t),
	\label{eq:fit_DCDW}
	\end{equation}
	
	which has been obtained using the edge-midpoint definition \cite{Maklar_2022,Harris_1996,Ding_1996} (see the SM for further details). The comparison between the gap dynamics and our model is reported in Fig.~\ref{fig:5}(c).

	\section{Discussion}
	
	We can now summarize the microscopic physical picture that stands at the base of the comparison between the dynamics obtained from the $3$TM and the ones extracted from the fits of the EDCs (Fig.~\ref{fig:5}(d)). Before the arrival of the pump pulse, the three subsystems given by electrons, hot phonons and lattice are in equilibrium, and their temperature is much lower than the CDW critical temperature: $T_{\text{e}}$ = $T_{\text{p}}$ = $T_{\text{l}}$ $<<$ $T_{\textsc{cdw}}$. Hence, the compound is characterized by a well-defined long-range CDW order. After the arrival of the visible pump pulse, the energy injected in the system is absorbed by the electrons and then transferred to the subset of strongly-coupled optical phonon modes (as expected from this picture, the dynamics does not depend on the pump photon energy, see the SM). The crucial point is that the excitation of these modes occurs without a macroscopic phase coherence, thus leading to a partial loss of the long-range order of the system \cite{Wall_2018}. This partial loss of the long-range order, that evolves on the same timescale of the population of the strongly-coupled phonon modes, entails a smearing and a partial closure of the CDW gap. Furthermore, additional simulations (reported in the SM) show that the dynamics of the CDW gap remains well described if any phonon energy between $19$ and $38\,$meV is considered. This result suggests a scenario in which not just a single phonon branch, but instead different branches, enclosed in a specific energy region, are involved in the disordering of the vanadium trimers, similarly to what was observed for the case of VO$_2$ \cite{Wall_2018, Cha_2023, Picano_2023, Erasmus_2012, Budai_2014}. After a first thermalization, electrons and hot phonons reach a common temperature and their dynamics become similar. Subsequently, the relaxation proceeds via anharmonic phonon-phonon interaction, which continues until a thermal equilibrium with all the other modes is reached. From here on, the relaxation is governed only by thermal diffusion, and hence it evolves on a much slower timescale, not captured by the $3$TM. Therefore, this microscopic picture also explains the dynamics observed at large pump-probe delays, \emph{i.e.} the presence of a plateau in both the gap closured and the gap broadening. At these delays the strongly coupled phonon modes have reached a common temperature with all the other modes, hence the relaxation process is brought forward on a much slower timescale solely by heat diffusion \cite{Block_2019}.

 	\section{Conclusions}
	
	In conclusion, we showed that the light-induced phase transition in VTe$_2$ can be fully described by considering the population of a subset of strongly-coupled optical phonon modes, with only a marginal role played by the excitation of the CDW amplitude modes. The excitation of the strongly-coupled phonons, which occurs without a macroscopic phase coherence, has the effect of increasing the lattice fluctuations and thus the transient disorder of the system \cite{Yang_2020, Cotret_2019}. The microscopic picture developed extends beyond the specific case of VTe$_2$ and could be applied to other CDW systems. More generally, the results presented in this article highlight the need for a deeper understanding of the interplay between lattice fluctuations and the CDW phase, paving the way for further exploration of the nonequilibrium properties of strongly-coupled CDW systems.
	
	\section{Acknowledgments}

	M.T. acknowledges financial support under the National Recovery and Resilience Plan (NRRP), Mission 4, Component 2, Investment 1.1, Call for tender No. 104 published on 2.2.2022 by the Italian MUR, funded by the European Union – NextGenerationEU– Project MEGS–CUP B53D23003930006. R.C acknowledges the Italian Ministry of Foreign Affairs and International Cooperation (MAECI), Grant no. PGR12320 - U-DYNAMEC - CUP B53C23006060001.  This work is performed in the framework of the Nanoscience Foundry and Fine Analysis (NFFA-MUR Italy Progetti Internazionali) facility.


\providecommand{\noopsort}[1]{}\providecommand{\singleletter}[1]{#1}%

\end{document}